\documentclass[nofootinbib,prd,twocolumn,showpacs,showkeys,preprintnumbers]{revtex4}
\usepackage{hyperref,amssymb,amsmath,mathrsfs,bm,graphicx}
\begin{document}
\title{Nature of the vorticity in the G\"{o}del spacetime}
\author{L. Herrera \footnote{Also at U.C.V., Caracas}} 
\email{laherrera@cantv.net.ve}
\affiliation{Departamento   de F\'{\i}sica Te\'orica e Historia de la  Ciencia,
Universidad del Pa\'{\i}s Vasco, Bilbao, Spain}
\author{A. Di Prisco} 
\email{alicia.diprisco@ciens.ucv.ve}
\affiliation{Departamento   de F\'{\i}sica, Facultad de   Ciencias, Universidad Central de Venezuela, Caracas, Venezuela}
\author{J. Ib\'a\~nez } 
\email{j.ibanez@ehu.es}
\affiliation{Departamento   de F\'{\i}sica Te\'orica e Historia de la  Ciencia,
Universidad del Pa\'{\i}s Vasco, Bilbao, Spain}
\date{\today}
\begin{abstract}
The physical meaning of the vorticity of the matter content in G\"{o}del spacetime is analyzed in some detail. As we shall see, unlike the situation in general stationary axially symmetric spacetimes (Lewis--Papapetrou), the vorticity in G\"{o}del spacetime is not associated to a circular flow of superenergy on the plane orthogonal to the vorticity vector. This fact might be at the origin of the strange behaviour of gyroscopes in such spacetime. The analysis emerging from the tilted version of   G\"{o}del spacetime supports further this point of  view. In order to tell apart the two situations (with and without circular flow of superenergy), we introduce two different  definitions of vorticity, related to the presence (absence) of such a flow.
\end{abstract}
\date{\today}
\maketitle
\section{INTRODUCTION}
The G\"{o}del spacetime \cite{goedel} is a stationary solution of the Einstein equations  with nonvanishing cosmological constant ($\Lambda$), whose matter content for comoving observers consists of dust  with constant density $\mu=-\frac{\Lambda}{4 \pi}=const.$ In this latter  expression, the appearing constant represents the only parameter of the solution.

It is well known that in such spacetime (for comoving observers) the congruence defined by the four velocity vector  is geodesic, shearfree, expansionfree but its vorticity is nonvanishing, i.e. matter rotates with respect to the compass of inertia  (see for example \cite{Adler}, \cite{Stephani}, \cite{Rindler}, \cite{chieh}, \cite{Barrow}, \cite{Gurses}).

Among the properties  of the G\"{o}del spacetime,  there are three particularly intriguing, namely:
\begin{itemize}
\item It admits closed timelike curves.
\item The energy--momentum tensor of this spacetime is exactly the same as the Einstein static universe \cite{Adler}.
\item The (coordinate) angular velocity of  particles moving on circular geodesics, as well as the precession of gyroscopes moving on circular geodesics, are independent on the parameter measuring the deviation of the space from flatness \cite{Rindler}.
\end{itemize}

The fact that the  precession of a  gyroscope moving on a geodesic circle is unaffected by the specific value of the parameter of the solution  is quite strange. Indeed  such a parameter appears (as it should be) in the expression for the curvature invariant. Therefore  the rate of precession would be independent on the magnitude of deviation from flatness, implying that it would be the same as in the flat spacetime limit!

For general axially symmetric stationary spacetimes (Lewis--Papapetrou) \cite{kramer} different parameters of the solution entering into the curvature invariants do affect the precession of gyroscopes moving on closed curves.

As we shall see below such a strange behaviour may be related to another property of  the vorticity in G\"{o}del spacetime, namely, it is not associated to a flow of superenergy on the plane orthogonal to the vorticity vector, as is the case for  stationary spacetimes of the Lewis --Papapetrou type \cite{H1}. 

To differentiate both situations we shall introduce the concepts  of ``dynamical  vorticity''  and ``kinematical vorticity'' to refer to situations where there is or there is not, respectively, superenergy flow on the plane orthogonal to the vorticity vector.

Finally, we shall consider the tilted version of  G\"{o}del spacetime, in this case there are additional terms for the vorticity, however there is not a component of superenergy on the plane orthogonal to the vorticity vector.

All these results are briefly summarized and commented in the last section.

\section{The G\"{o}del spacetime}
We shall closely follow (with slight changes) the notation in \cite{Rindler}, thus for the line element we have (relativistic units, $G=c=1$ are used throughout the paper):

\begin{equation}
ds^2=-4R^2\left\{(dt+\sqrt{2} S^2d\phi)^2-[dr^2+(S^2+S^4) d\phi^2+dz^2]\right\},
\label{1}
\end{equation}
where $S\equiv \sinh{r}$, $R$ is a constant, and we number coordinates $x^0=t,x^1=r,x^2=\phi,x^3=z$.

Such  a metric satisfies Einstein equations with cosmological constant and 
\begin{equation}
T_{ab}=\mu v_a v_b.
\label{4}
\end{equation}
For a comoving observer
\begin{equation}
v^a=(\frac {1}{2R},0,0,0),\label{5}
\end{equation}
and the following identity holds
\begin{equation}
\mu=\frac{1}{8\pi R^2}=-\frac{\Lambda}{4\pi}.\label{6}
\end{equation}

For the congruence defined by (\ref{5}) the expansion scalar, the four--acceleration and the shear tensor vanish, however the vorticity does not.
Indeed, one obtains for the vorticity vector
\begin{equation}
\omega_i=\frac{1}{2}\eta_{ijkl}v_{m;n}v^lg^{jm}g^{kn}=\left(0,0,0,\sqrt {2}\right),\label{7}
\end{equation}
or
\begin{equation}
\omega^i=\left(0,0,0,\frac{1}{2R^2\sqrt{2}}\right),\label{8}
 \end{equation}
producing

\begin{equation}
\omega^2=\omega_i\omega^i=\frac {1}{2R^2 },\label{9}
\end{equation}
where $\eta_{ijkl}$ is the Levi--Civita tensor.

It is remarkable that even though $R$ appears in the expression for $\omega$, it does not affect the precession of a gyroscope moving along a circular geodesic. Indeed the change of orientation of such a gyroscope with respect to the original   G\"{o}del lattice is (see \cite{Rindler} for details):
\begin{equation}
\Delta \phi=2\pi-\pi\sqrt{1-\sinh^2{2r}}.
\label{10}
\end{equation}
It should be stressed that $R$ nor does affect the precession with respect to a second gyroscope fixed in the original  G\"{o}del lattice \cite{Rindler}.

The above result becomes more intriguing if we observe that for the Riemann invariant we obtain 
\begin{equation}
I=R_{ijkl}R_{mnrs}g^{im}g^{jn}g^{kr}g^{ls}=\frac{3}{R^4},\label{11}
\end{equation}
clearly indicating that the parameter $R$  measures deviations from flatness.

Thus expression (\ref{10}) determines  the  value for the precession of the gyroscope independently on the magnitude of the curvature of the spacetime. This  situation clearly differs from the case of the  Kerr  spacetime.

In order to delve deeper into this point  it will be useful to calculate the super--Poynting vector  \cite{11}, \cite{12}, \cite{13}, \cite{14}, defined by 
\begin{equation}
P_i=\eta_{jikl}v^j(Y_{mn}Z^{kn}-X_{mn}Z^{nk})g^{lm},\label{12}
\end{equation}
where $Y_{ij}$ (the electric part of the Riemann tensor), $Z_{ij}$ (the magnetic part of the Riemann tensor) and $X_{ij}$ are defined by:
\begin{eqnarray}
Y_{ij}&=&R_{ikjl}v^k v^l=\nonumber\\
&=& \left[ 
\begin {array}{cccc} 0&0&0&0\\ \noalign{\medskip}0&2&0&0
\\ \noalign{\medskip}0&0&2 \sinh^2 {r} \cosh^2 {r}&0
\\ \noalign{\medskip}0&0&0&0\end {array} \right]
\end{eqnarray}'\label{13}
\begin{equation}
Z_{ij}=R^*_{jkil} v^k v^l=0
\label{14}
\end{equation}

\begin{equation}
X_{ij}=^*R^*_{ikjl}v^k v^l=0,\label{15}
\end{equation}
with 
\begin{eqnarray}
R^*_{ijkl}&=&\frac{1}{2}\eta_{mnkl}R_{ijps}g^{pm}g^{sn},\\\label{16}
^*R^*_{ijkl}&=&\frac{1}{2}\eta_{ijmn}R^*_{pskl}g^{mp}g^{ns}.\label{17}
\end{eqnarray}
From the above it is obvious that  $P_i=0$. It should be observed that the three tensors $Y, Z, X$ may also be obtained from the Weyl tensor, and from them another super-Poynting vector can be obtained. In vacuum both expressions coincide of course, but in matter they differ. However it is a simple matter to check that the magnetic part of the Weyl tensor  vanishes too, and therefore so does the ensuing super--Poynting vector.

Now, the interest of the above result stems from the fact that for stationary (Lewis--Papapetrou) spacetimes (e.g. Kerr) the association of  vorticity (and its resulting effects) and the existence  of a circular flow of superenergy on the plane orthogonal to the vorticity vector has been established (see \cite{H1} for details).

The facts exhibited above suggest the existence of  two  ``kinds''  of  vorticity. On the one hand we have a vorticity always associated to the existence of  a circular flow of superenergy on the plane orthogonal to the vorticity vector and affecting the precession of a gyroscope in a way that is dependent on the essential parameters of the metric describing the spacetime under consideration (by essential we mean those parameters that cannot be removed by any coordinate transformation, and enter into the expression of curvature invariants). We shall call this kind of vorticity dynamical vorticity, one example of which  is provided by the Kerr metric.

On the other hand  we shall refer to kinematical vorticity, whenever  the latter  neither is associated to a circular flow of superenergy on the plane orthogonal to the vorticity vector, nor is the precession of a gyroscope affected by the essential parameters of the metric. The G\"{o}del  spacetime provide a good example of this kind of vorticity.

In order to delve deeper into this issue, we shall next consider  the tilted version of the G\"{o}del spacetime.
\section{Tilted G\"{o}del spacetime}
As   is well known there exists in general relativity a certain degree of arbitrariness in the choice of the four velocity vector in terms of which  the energy--momentum tensor is split, which leads to a variety of different physical interpretations of the source of a given spacetime. Such arbitrariness is in  its turn related to the choice of  the congruence of observers (see \cite{t1}-\cite{t11} and references therein).

The description  of the G\"{o}del  spacetime given in the previous section corresponds to the congruence of observers at rest with respect to the dust distribution. In order to obtain the tilted version, of the  G\"{o}del  spacetime given by (\ref{1}), we have to obtain the tilted congruence. For doing that  we  have to perform a Lorentz boost from the locally comoving Minkowskian frame (associated to $v^\mu$) to the locally Minkowskian frame with respect to which any fluid element has velocity $u$ in the $r$ direction. For simplicity we shall consider a boost only in the ``radial'' direction and $u$ to be a function of $t$ and $r$ alone.

Then, the corresponding tilted congruence is characterized by the four--velocity vector
\begin{equation}
\tilde v^a=\left(\frac {1}{2R\sqrt {1-{
u}^{2}}},\frac {u}{2R\sqrt {1-{u}^{2}}},0,0\right),\label{t1}
\end{equation}
from which all the kinematical quantities can be calculated. They will not be displayed here since we shall not use them. Suffice is to say that now the fluid is nongeodesic, shearing, expanding (it is worth noticing that shearing and expanding versions of  the G\"{o}del  spacetime may also be obtained by perturbation \cite{Barrow}) and  the vorticity  vector and vorticity scalar  read
\begin{equation}
\tilde \omega^i=\left(0,0,0,-\frac {1}{4\sqrt{2}}\frac{\dot u\sinh r-2\cosh r}{R^2(1-u^2)\cosh r}\right),\label{tn}
 \end{equation}

\begin{equation}
\tilde  \omega^2=\frac{(\dot u\sinh r-2\cosh r)^2}{8R^2(1-u^2)^2\cosh^2 r},\label{t2}
\end{equation}
exhibiting the contribution  of the tilting velocity to the vorticity and reducing to (\ref{8}) and (\ref{9}) in the non--tilted case (overdot denotes derivative with respect to $t$).

For the tilted observer the matter distribution is  no longer dust but a dissipative anisotropic fluid, whose energy momentum tensor is
\begin{equation}
T_{ij}=\tilde \mu \tilde v_i \tilde v_j+P h_{ij}+\Pi_{ij}+\tilde q_i \tilde v_j+\tilde q_j \tilde v_i \label{t3}
\end{equation}
where $\tilde \mu, P, \Pi_{ij}, \tilde q_i $ denote the energy density, the isotropic pressure, the anisotropic pressure tensor and the heat flux vector respectively. From  the Einstein equations it follows that (we omit the expression for $\Pi_{ij}$ since it is quite cumbersome and we do not need it here)
\begin{equation}
\tilde \mu=\frac{1}{R^2(1-u^2)},
\end{equation} \label{t4}

\begin{equation}
P=\frac{1}{3}\,\frac{u^2}{R^2(1-u^2)} \label{t5}
\end{equation}

\begin{equation}
\tilde q^i= \left[-\frac {u^{2}}{{2R}^{3}\left( 1-{u}^{2} \right) ^{3/2}},-\frac {u}{{2R}
^{3} \left( 1-{u}^{2} \right) ^{3/2}},0,0\right]. \label{t6}
\end{equation}

Nex, the calculations of $(X, Y, Z)$ tensors   yields:
\begin{widetext}
\begin{eqnarray}
\tilde Y_{ij}&=&R_{ikjl}\tilde v^k \tilde v^l=\nonumber\\
&=& \left[ 
\begin {array}{cccc} \frac {{2u}^{2}}{1-{u}^{2}}&\frac {-2u}{
1-{u}^{2}}&\frac {2\sqrt {2} u^{2} \sinh^2r}{1-{u}^{2}}&0\\ \noalign{\medskip}-\frac {2u}
{1-{u}^{2}}&\frac{2}{1-{u}^{2}} &-\frac {2\sqrt {2} u\sinh^2 r}{1-{u}^{2}}&0
\\ \noalign{\medskip}\frac {2\sqrt {2} u^{2}\sinh^2 r}{1-{u}^{2}}&-\frac {2\sqrt {2}u\sinh^2 r}{1-{u}^{2}}&
\frac { 2\sinh^2r \left( 
\cosh ^2r  +3 {u}^{2}  \cosh^2  r  -2{u}^{2} \right) }{1-{u}^{2}}&0
\\ \noalign{\medskip}0&0&0&0\end {array} \right],
\end{eqnarray}  \label{t7}
\end{widetext}
\begin{equation}
\tilde Z_{ij}=R^*_{jkil} v^k v^l=\left[ 
\begin {array}{cccc} 0&0&0&0\\ \noalign{\medskip}0&0&0&0
\\ \noalign{\medskip}0&0&0&0\\ \noalign{\medskip}0&0&-\frac {4u\sinh r \cosh  r } {1-{u}^{2}}&0
\end {array} \right], \label{t8}
\end{equation}

\begin{equation}
\tilde X_{ij}=^*R^*_{ikjl}v^k v^l=\left[ 
\begin {array}{cccc} 0&0&0&0\\ \noalign{\medskip}0&0&0&0
\\ \noalign{\medskip}0&0&0&0\\ \noalign{\medskip}0&0&0&{\frac {2(1+{
u}^{2})}{1-{u}^{2}}}\end {array} \right]. \label{t9}
\end{equation}

From the above expressions  the super-Poynting vector calculated for the tilted congruence produces

\begin{equation}
\tilde P^i=\left[-\frac {
 \left( 1+{u}^{2} \right) {u}^{2}}{2{R}^{5} \left( 1-{u}^{2} \right) ^{
5/2}},-\frac { \left( 1+{u}^{2} \right) u}{{2R}^{5} \left( 1-{u}
^{2} \right) ^{5/2}},0,0\right] .\label{t10}
\end{equation}

Observe that now the super-Poynting vector does not vanish, however it has only ``radial'' component (besides the timelike component)  and therefore there  is  no circular flow on the plane orthogonal to the vorticity vector. In fact such non--vanishing super--Poynting vector is related to the ``radial'' heat flux vector, as it happens in the spherically symmetric case (see \cite{t12} for  a discussion on this point).
It is also worth mentioning that even though the magnetic part of the Weyl tensor does not vanish for the tilted congruence, the  ensuing super--Poynting vector does. Thus the extra terms appearing  in the vorticity of the tilted congruence  (following the  terminology introduced in the previous section)  are also of ``kinematical'' nature.

\section{SUMMARY}
We have established that  the vorticity in the  G\"{o}del spacetime is not related to the presence of a circular flow of super--energy on the plane orthogonal to the vorticity vector. The absence of such a flow, which is always present in Lewis--Papapetrou stationary metrics, together with the fact that the  curvature does not affect the precession of a gyroscope in the  G\"{o}del spacetime, suggests that  there are two different classes of vorticity. We call ``dynamical'' vorticity, the rotation of the lattice relative to the compass of inertia, that is always accompanied of  a circular flow  of superenergy on the plane orthogonal to the vorticity vector. When such a flow is absent we talk about ``kinematical'' vorticity. This is the case of the  vorticity in the  G\"{o}del spacetime.

Next we analyzed the tilted version of the  G\"{o}del spacetime. In this case some extra terms appear in the vorticity of the tilted congruence and the super--Poynting vector (constructed from the Riemann tensor) is not vanishing. However such a vector  has no components on the plane orthogonal to the vorticity vector and the radial nonvanishing component is related to the  heat flux observed by the tilted observer. In other words such a vorticity is also ``kinematical''. It is worth mentioning that this is also the case for the tilted Szekeres spacetime.  Indeed, in the standard (non--tilted version) \cite{1} \cite{2}, the congruence defined by the four--velocity has vanishing vorticity, whereas the tilted observers detect vorticity in the congruence of fluid world lines \cite{t11}. However neither in this case there is a component of the super--Poynting vector on the plane orthogonal to the vorticity vector and the nonvanishing radial component is associated to the heat flux vector detected by the tilted observer. Thus the vorticity observed by tilted observers in Szekeres spacetime is also kinematical.

We conclude with  the  following comment: As we mentioned in the Introduction, the  G\"{o}del spacetime admits closed timelike curves, whereas other spacetimes with vorticity, do not. Thus the question arises about the possibility that closed timelike curves are specifically associated to  ``kinematical'' vorticity. Although we do not answer here to the above question, we believe that this issue deserves further attention.

\begin{acknowledgments}
This work was partially supported by the Spanish Ministry of Science and Innovation (Grant FIS2010-15492). 
\end{acknowledgments} 

 \end{document}